\documentclass{article}
\usepackage{moreverb,bm,xspace,subfig,inconsolata}
\usepackage[colorlinks,bookmarksopen,bookmarksnumbered,citecolor=red,urlcolor=red]{hyperref}
\usepackage{graphicx}
\usepackage{amsmath}

\newcommand\BibTeX{{\rmfamily B\kern-.05em \textsc{i\kern-.025em b}\kern-.08em
T\kern-.1667em\lower.7ex\hbox{E}\kern-.125emX}}

\newcommand{\ie}{{\em i.e.\/}\xspace}
\newcommand{\eg}{{\em e.g.\/}\xspace}

\usepackage{lineno}

\begin{document}

\title{Estimating the Expected Value of Partial Perfect Information in Health Economic Evaluations using Integrated Nested Laplace Approximation}
\author{Anna Heath, Ioanna Manolopoulou, Gianluca Baio}
\date{Department of Statistical Science, University College London, Department of Statistical Science, University College London (UK)}

\maketitle
\begin{abstract}
The Expected Value of Perfect Partial Information (EVPPI) is a decision-theoretic measure of the ``cost'' of parametric uncertainty in decision making used principally in health economic decision making. Despite this decision-theoretic grounding, the uptake of EVPPI calculations in practice has been slow. This is in part due to the prohibitive computational time required to estimate the EVPPI via Monte Carlo simulations. However, recent developments have demonstrated that the EVPPI can be estimated by non-parametric regression methods, which have significantly decreased the computation time required to approximate the EVPPI. Under certain circumstances, high-dimensional Gaussian Process regression is suggested, but this can still be prohibitively expensive. Applying fast computation methods developed in spatial statistics using Integrated Nested Laplace Approximations (INLA) and projecting from a high-dimensional into a low-dimensional input space allows us to decrease the computation time for fitting these high-dimensional Gaussian Processes, often substantially. We demonstrate that the EVPPI calculated using our method for Gaussian Process regression is in line with the standard Gaussian Process regression method and that despite the apparent methodological complexity of this new method, \verb"R" functions are available in the package \verb"BCEA" to implement it simply and efficiently.
\end{abstract}

\section{Introduction}
Broadly speaking, the objective of publicly funded health care systems, such as the UK National Health Service, is to maximise health gains across the general population, given finite monetary resources and limited budget. Bodies such as the National Institute for Health and Care Excellence (NICE) in the UK provide guidance on decision-making on the basis of health economic evaluation. This covers a suite of analytic approaches for combining costs and clinical consequences of an intervention, in comparison to alternative options which may already be available, with the aim of aiding decision-making associated with health resources. Much of the recent research has been oriented towards building the health economic evaluation on sound and advanced statistical decision-theoretic foundations, arguably making it a branch of applied statistics \cite{Briggsetal:2006,WillanBriggs:2006} and increasingly often under a Bayesian approach \cite{O'HaganStevens:2001,O'Haganetal:2001,Spiegelhalteretal:2004,Baio:2012}.

In a nutshell, the process involves the identification of suitable measures of clinical benefits (generically termed as ``effectiveness'') and costs associated with an intervention, which we indicate as $(e,c)$. The variable $c$ usually includes the cost of acquisition and implementation of the health intervention (e.g.\ a drug), or societal costs such as those related to number of days off work or social care. As for the clinical benefits $e$, they can be a ``hard'' measurement (e.g.\ number of cases averted), but most often are considered in terms of \textit{Quality Adjusted Life Years} (QALYs) \cite{LoomesMcKenzie:1989}, combining the quantity and the quality of life provided by a given intervention. Individual level variability in the outcome is expressed in terms of a joint probability distribution $p(e,c\mid\bm\theta)$, indexed by a set of parameters~$\bm\theta$ whose uncertainty is described by a probability distribution $p(\bm\theta)$, in a Bayesian context.

According to the precepts of decision theory \cite{RaiffaSchlaifer:1961}, for each intervention $t=0,\ldots,T$ ($t=0$ being the ``reference'' alternative, e.g.\ standard of care and $t=1,\ldots,T$ the new options being evaluated) the health economic outcomes $(e,c)$ are combined in order to quantify the overall ``value'' of the intervention, usually in terms of the  monetary net benefit which is given by the costs associated with the treatments subtracted from rescaled effectiveness (we define this formally in \S\ref{general_framework}). The alternative associated with the highest expected net benefit is deemed as ``the most cost-effective'', given current evidence --- notice that in a Bayesian context, this expectation is taken over the distributions of both the individual level outcomes and population level parameters. From the decision-theoretic point of view, the identification of the overall expected net benefit is all that is needed to reach the optimal decision given the current state of knowledge available to the decision-maker \cite{Claxton:1999b,Claxtonetal:2000}.

However, the implementation of a health care intervention is usually associated with risks such as the irreversibility of investments \cite{Claxton:1999}. Moreover, health economic models often involve a relatively large number of parameters, usually estimated using limited information. For these reasons, health technology assessment (HTA) bodies such as NICE recommend a thorough investigation of the impact of uncertainty on the decision making process of parametric and model uncertainty, a process known in the health economics literature as \textit{Probabilistic Sensitivity Analysis} (PSA) \cite{Claxtonetal:2005,eunethta:2014,CADTH:2006,Australia:2008}. 

The analysis of the \textit{value of information} (VoI) \cite{Howard:1966} is an increasingly popular method to conduct PSA in health economic evaluations \cite{FelliHazen:1998,FelliHazen:1999,Claxton:1999,Claxtonetal:2001,Adesetal:2004,BrennanKharroubi:2005,Briggsetal:2006,Fenwicketal:2006}. The basic idea of VoI analysis is to compare the decision based on current evidence to the one that would be made, had the uncertainty in the parameters been resolved e.g.\ by observing an (infinitely) large amount of data. 

The main advantage of the analysis of the VoI is that it directly addresses the potential implications of current uncertainty, not only in terms of the likelihood of modifying the current decision in light of new and more definitive evidence, but also in terms of the opportunity cost of the incorrect decision. If this cost is low, there is little value in agonising about the decision, even for a low probability of cost-effectiveness, as the implicit penalty is negligible if the decision turns to be wrong in the face of new evidence. Therefore, the probability of cost-effectiveness alone can dramatically overstate or down-play the decision sensitivity. For this reason, it has been advocated that VoI measures should also be presented when representing decision uncertainty \cite{FelliHazen:1998,Baio:2012,Meltzer:2001,Campbelletal:2014}.

Despite the useful features of a VoI analysis, its uptake in health economic evaluation has been slow. VoI analysis has been hindered by several different factors, both theoretical and practical. Theoretically, issues such as whether the economic model can be trusted and what constitutes a ``high'' level of decision uncertainty given that VoI measures are unbounded. Practically, VoI measurements can be computationally costly unless restrictive assumptions are made \cite{Wilson:2015}. It is typically easy to calculate numerically (but not analytically) the expected value of learning \textit{all} the model parameters perfectly. This is known as the (overall) Expected Value of Perfect Information (EVPI). However, this quantity has often little practical use, as it will be rarely possible to learn all the model parameters at once in a new study. Thus, the decision-maker is usually interested in the expected value of information about subsets of parameters of interest, sometimes called focal parameters. This will indicate which parameters are driving decision uncertainty, and thus where future research may add value. This subset analysis is concerned with the computation of the \textit{Expected Value of Perfect Partial Information} (EVPPI) and is usually computationally~costly. 

In the last few years, research has focused on alternative methods that could be used to speed up the estimation of the EVPPI without compromising its accuracy, so as to increase its applicability in health economic evaluations \cite{Madanetal:2014,StrongOakley:2013,Sadatsafavietal:2013}. A review of these methods is given in \cite{CoyleOakley:2008}. A promising development \cite{StrongOakley:2014} has recently explored the use of non-parametric regression methods, specifically Generalised Additive Models (GAMs) \cite{Hastie:1990} and Gaussian Process (GP) regression \cite{Rasmussen:2006}, to approximate the EVPPI. These methods have been shown to be very effective: GAMs are very flexible and extremely inexpensive in computational terms if the number of  parameters of interest $P$ is relatively small (e.g.\ $P<5$). When $P$ is large, however, GAMs either overfit or simply cannot be used as the number of parameters exceeds the number of data points. GP regression methods may be used as an alternative regression method: on the one hand, they overcome the limitations of GAMs and can be used to estimate the EVPPI for higher dimensions of the subset of parameters of interest, producing a significant improvement over simpler, but computationally prohibitive methods. On the other hand, however, the computational cost to fit a GP regression model to a realistic problem including a relatively large number of parameters can still be substantial. Moreover, as EVPPI values may need to be calculated for a large number of different parameter sets, this computational time can be multiplied 10- or 100-fold.

To overcome this issue, we propose in this paper a faster method to fit GP regression, based on spatial statistics and Bayesian inference using Integrated Nested Laplace Approximation \cite{Rueetal:2009}. We translate the estimation of the EVPPI into a ``spatial'' problem by considering that the simulated net benefit values, \ie the values from the PSA, are ``observed'' at different points in the parameter space. We can therefore use the available technology for fast Bayesian computation of spatial models \cite{Lindgren:2012,Blangiardo:2013} to approximate the EVPPI efficiently. Furthermore, as this spatial machinery loses its computational advantages in higher dimensional spaces, we demonstrate that the use of dimension-reduction techniques to project onto a 2-dimensional space is accurate for the examples considered.  Thus, we can use this method along with dimensionality reduction to approximate the EVPPI quickly and efficiently in applied cases, irrespective of the complexity of the problem. 

The paper is structured as follows: in \S \ref{general_framework} we present the general framework used for the analysis of the value of information in health economic evaluation and in \S \ref{GP} we briefly review the main characteristics of GPs and specifically their application in the computation of the EVPPI. Then in \S \ref{our_stuff} we present our proposal for a new method to compute the EVPPI; first we briefly review the main features of the spatial statistics literature based on stochastic partial differential equations (described in \S \ref{SPDE}), which is used to estimate the correlation function required to fit the GP. Then, in \S \ref{method} we discuss how this can be brought to bear in the modelling and efficient computation of the EVPPI. In \S \ref{results} we test our method in comparison to existing GP regression models to estimate the EVPPI on a set of health economic examples. We particularly focus on the issues of computational time as well as accuracy of the estimation. Finally, in \S \ref{tech} and \S \ref{discussion} we present some technical aspects as well as the main conclusions from our work.

\section{Value of information analysis in health economics}\label{general_framework}
PSA is usually based on a simulation approach \cite{BaioDawid:2011,Baio:2012,Andronisetal:2009}: uncertainty about the relevant parameters $\bm\theta$ is described by a suitable probability distribution, from which a sample of $S$ values is obtained, e.g.\ via Monte Carlo (MC) sampling from the prior or Markov Chain MC (MCMC) estimation of the posteriors under a Bayesian framework, or using bootstrap in a frequentist approach. First, for each intervention, the expected value is computed conditionally on each value of the simulated parameters. Assuming the commonly used \textit{monetary net benefit} \cite{StinnettMullahy:1998} to quantify the value of the different treatments, this value is estimated by
\begin{eqnarray*}
\mbox{NB}_t(\bm\theta_s) = k\mbox{E}[e\mid \bm\theta_s; t]-\mbox{E}[c\mid\bm\theta_s; t],
\end{eqnarray*}
where $\bm\theta_s$ is the $s$-th set of simulated values for the parameters vector, $e$ and $c$ are the measures of effectiveness and cost respectively and $k$ is the \textit{willingness to pay}, which is used to put the cost and effectiveness measures on the same scale, \ie in terms of the amount of money that the decision maker is willing to pay to increment the benefits by one unit. Notice here that $(e,c)$ represent the individual level effectiveness and costs for a specific value of $\bm\theta$ and conditional on any observed data. Therefore, the expectation is taken over the joint distribution of $(e,c)$ to give the population level net benefit for the specific value of $\bm\theta$.

The vector of values $\mbox{\textbf{NB}}_t=\left[\mbox{NB}_t(\bm\theta_1),\ldots,\mbox{NB}_t(\bm\theta_S)\right]'$ is a sample from the distribution of the decisions (randomness being induced by uncertainty in the parameters) and can be analysed to determine the impact of parameter uncertainty on the decision-making process. If the optimal decision, \ie the intervention with the maximum expected net benefit, varies substantially across the simulations, then the decision-making process is sensitive to the uncertainty in the model parameters and more research could be recommended by the HTA bodies. 

For example, consider a non life-threatening infectious disease (such as mild influenza) and assume that under current practice ($t=0$), individuals have a risk of infection $\pi$. If they become infected, they are subject to an average duration of the disease of $\lambda$ days, for each of which they have to follow a course of treatment that costs $\gamma$ monetary units (say, \pounds). The new treatment, whose cost-effectiveness is being assessed ($t=1$), has an average implementation cost of \pounds $\xi$ and it is assumed that the chance of infection is reduced by a factor $\rho$. However, if an individual becomes infected then they will still have a period of $\lambda$ days in which they will require treatment at the cost of \pounds $\gamma$ per day. Under these (unrealistically!) simple assumptions, we can define $\bm\theta=(\pi,\lambda,\gamma,\xi,\rho)$. Assuming further that suitable probability distributions can be associated with each of the elements of $\bm\theta$ to describe our current uncertainty, we could reasonably define the net benefits for the two interventions as 
\[ \mbox{NB}_0(\bm\theta) = k(-\pi\lambda) - \pi\gamma\lambda \qquad \mbox{ and } \qquad \mbox{NB}_1(\bm\theta) = k(-\pi\rho\lambda) - (\xi+\pi\rho\gamma\lambda), \]
implying that the clinical benefit is given by minus the chance of being infected multiplied by the length of the infection and that cost is given by the sum of the implementation cost of the intervention and the overall cost of treatment if infected.

In general terms, the expected opportunity loss of making a decision based on current evidence instead of on perfect information can be quantified by the \textit{Expected Value of Perfect Information}, defined as
\begin{equation} 
\mbox{EVPI} =  \mbox{E}_{\bm\theta}\left[\max_t\mbox{NB}_t\left(\bm\theta\right)\right] - \max_t\mbox{E}_{\bm\theta} \left[\mbox{NB}_t\left(\bm\theta\right)\right],\label{EVPI-eq}
\end{equation} 
where the expectation is taken with respect to the posterior distribution of $\bm\theta$.

Since these expectations are typically not analytically available, they are estimated through simulations. Provided the number $S$ of simulations used to perform PSA is large enough to characterise the underlying distribution of the decisions, it is straightforward to compute a MC estimate using the simulated values for the net benefits
\[ \widehat{\mbox{EVPI}} = \frac{1}{S}\sum_{s=1}^S \max_t \mbox{NB}_t(\bm\theta_s) - \max_t \frac{1}{S} \sum_{s=1}^S \mbox{NB}_t(\bm\theta_s), \]
which usually requires almost no extra computational time, once the PSA samples are available.

In most practical situations, however, the interest is in quantifying the value of reducing uncertainty on a specific subset $\bm\phi \subset \bm\theta$ of parameters of interest as it may be feasible to conduct a specific clinical trial or literature review in order to potentially reduce the level of current uncertainty in these parameters. For example, for the simple model described above we may be interested in learning the value of investigating the epidemiological parameters $\bm\phi=(\pi,\lambda,\rho)$ describing the risk and duration of the infection and the effectiveness of the intervention in reducing it, while considering the remaining parameters $\bm\psi=(\gamma,\xi)$ as nuisance.

This value is known as the Expected Value of Perfect \emph{Partial} Information and in the general setting is defined as
\begin{equation}
\mbox{EVPPI} = \mbox{E}_{\bm\phi} \left[\max_t \mbox{E}_{\bm\psi\mid\bm\phi} \left[\mbox{NB}_t(\bm\phi,\bm\psi) \right]\right]- \max_t \mbox{E}_{\bm\theta}\left[ \mbox{NB}_t (\bm\phi,\bm\psi)\right],\label{EVPPI-eq}
\end{equation}
where $\bm\theta=(\bm\phi,\bm\psi)$. The last term in equation (\ref{EVPPI-eq}) is again the maximum expected net benefit under current evidence. The first term is made by two nested elements: the inner part is the maximum expected net benefit that would be obtained if uncertainty in the parameters of interest \textit{only} were resolved. This means that the inner expectation assumes that the value of $\bm\phi$ is known. Of course, as the ``true'' value of $\bm\phi$ is not known, it is necessary to consider the expectation over the current distribution of $\bm\phi$.

In simple cases, where the conditional expectation $\mbox{E}_{\bm\psi\mid\bm\phi} \left[\mbox{NB}_t(\bm\phi,\bm\psi) \right]$ is available analytically as a function of $\bm\phi$, then it is possible to calculate the EVPPI using a single sample from $p(\bm\phi)$. This can occur in several settings, the most common of which is known as the ``sum-product'' form for the net benefit. This allows us to calculate the EVPPI based on samples that have already been obtained for PSA as part of a standard health economic analysis.

A more general solution, which involves additional sampling, is to use a nested MC scheme, in which first a sample $\bm\phi_1,\ldots,\bm\phi_{S_\phi}$ is obtained from the marginal distribution of $\bm\phi$ and then, for each $s=1,\ldots,S_\phi$, a sample $\bm\psi_1,\ldots,\bm\psi_{S_\psi}$ from the conditional distribution $p(\bm\psi\mid\bm\phi_s)$ is also simulated. This produces a total of $S_\phi \times S_\psi$ simulations, where both numbers need to be large in order to reduce the Monte Carlo error such that the EVPPI estimates are suitably precise; for example, Brennan et al.\ \cite{Brennanetal:2007} suggest that $S_\phi$ and $S_\psi$ should be in the order of 10\,000, although this may need to be higher in complex models. This immense computational burden and the difficulty in deriving analytic results in practical scenarios have been arguably the main reasons for the relatively limited practical use of the EVPPI as a tool for PSA \cite{Steutenetal:2013,Tuffahaetal:2014}. 

\subsection{Gaussian Process regression for efficient computation of the EVPPI} \label{GP}
Gaussian Processes are a family of stochastic processes used in statistics and machine learning for non-parametric regression, classification and prediction \cite{Rasmussen:2006,Davis:2001} and can be thought of as an extension of the multivariate Normal distribution to an infinite vector of observations \cite{Ebden:2008,Rasmussen:2006}. Strictly speaking, a GP is an infinite collection of random variables, any subset of which follows a multivariate Gaussian distribution \cite{Rasmussen:2004}. A GP is entirely defined in terms of its mean and covariance \textit{functions} \cite{Mackay:1998,GibbsMackay:1997}, which calculate the mean vector and covariance matrix for each subset of random variables depending on some input values and a small set of hyperparameters. These inputs determine the specific mean and variance for each random variable in the process. Consequently, GPs can be used for regressing random variables on a set of input~values.

To fit a GP for non-parametric regression, the general form of the mean and covariance function is specified by the modeller. In general, the covariance function is a taken as a decreasing function of the ``distance'' between any two input values, \ie points that are ``closer'' have a higher correlation \cite{Rasmussen:2006, OHagan:1978} where ``distance'' and ``closeness'' can be measured in different ways depending on the context. These functions typically depend on a set of hyperparameters; for example, the covariance function is often defined in terms of a smoothness parameter that determines the similarity between two points ``close'' together and a GP marginal variance parameter. Once these general functions are specified, problem-specific values for the hyperparameters need to be determined.

In a Bayesian setting, vague and conjugate priors have been proposed for the hyperparameters allowing for partially analytically tractable posterior distributions \cite{StrongOakley:2014,OHagan:1991}. Integration and numerical optimisation can then be used to find maximum a posteriori estimates of the GP parameters. Therefore, GPs are an increasingly popular method of regression since their extreme flexibility typically is obtained at a relatively small computational cost as MCMC methods may be avoided to fit them. However, for large datasets the cost of fitting a GP is still substantial as numerical optimisation in this setting requires inverting an $S\times S$ dense matrix, at a computational cost of $\mathcal{O}(S^3)$.

The basic idea exploited by Strong et al.\ \cite{StrongOakley:2014} is to consider the net benefit of each treatment $t$ computed using the $s-$th set of simulated values of the parameters as a noisy observation of the ``true'' underlying conditional expectation
\begin{eqnarray}
\mbox{NB}_t(\bm\theta_s) = \mbox{E}_{\bm\psi\mid\bm\phi_s}\left[ \mbox{NB}_t(\bm\theta) \right] +\varepsilon_s, \label{NB_GP1} 
\end{eqnarray} 
with $\varepsilon_s \stackrel{\rm{iid}}{\sim} \mbox{Normal}(0,\sigma^2_\varepsilon)$ and assuming conditional independence between the net benefits under the different treatments~$t$. As the conditional expectation on the right hand side changes as a function of the  parameters of interest only, we can equivalently write (\ref{NB_GP1}) as
\begin{eqnarray}
\mbox{NB}_t(\bm\theta_s) = g_t(\bm\phi_s)+ \varepsilon_s.  \label{NB_GP} 
\end{eqnarray}
Once the functions $g_t(\cdot)$ have been estimated using GP regression methods, the fitted values $\hat{g_t}(\bm\phi_s)$ can be used to approximate the EVPPI by computing
\begin{equation*}
\widehat{\mbox{EVPPI}} = \frac{1}{S} \sum_{s=1}^S \max_t \hat{g_t}(\bm\phi_s) - \max_t \frac{1}{S}\sum_{s=1}^S \hat{g_t}(\bm\phi_s). 
\end{equation*}

Assuming a GP structure for the functions $g_t(\cdot)$ in a linear regression framework effectively amounts to modelling
\begin{equation}
\left(\begin{array}{c}
g_t(\bm\phi_1)\\
g_t(\bm\phi_2)\\
\vdots\\
g_t(\bm\phi_S)
\end{array}\right) \sim \mbox{Normal}(\bm H \bm\beta, \bm\Sigma),\label{g_GP}
\end{equation}
where: $\bm\phi_s$ is the $s$-th simulated value for $\bm\phi$; $\bm H$ is a design matrix 
\begin{equation} 
\bm H =\left(\begin{array}{c c c c}
1 & \phi_{11} & \cdots & \phi_{1P}\\
1 & \phi_{21} & \cdots & \phi_{2P}\\
\vdots&  & \ddots\\
1 & \phi_{S1} & \cdots & \phi_{SP}\\
\end{array}\right); \label{H}
\end{equation} 
$\bm\beta$ is the vector of regression coefficients describing the linear relationship between the parameters of interest $\bm\phi$ and the conditional expectation of the net benefits; and the covariance matrix $\bm\Sigma$ is determined by the covariance function $\bm{\mathcal{C}}$, a matrix operator whose elements $\mathcal{C}(r,s)$ describe the covariance between any two points $g_t(\bm\phi_r)$ and $g_t(\bm\phi_s$). 

Strong et al.\ \cite{StrongOakley:2014} use a squared exponential, also known as an exponentiated quadratic, covariance function $\bm{\mathcal{C}}_{\rm{Exp}}$, defined by
\begin{equation}
\mathcal{C}_{\rm{Exp}}(r,s) = \sigma^2 \exp\left[-\sum_{p=1}^{P} \left(\frac{\phi_{rp}-\phi_{sp}}{\delta_p}\right)^2\right] \label{ExpCov}
\end{equation} 
where $\phi_{rp}$ and $\phi_{sp}$ are the $r$-th and the $s$-th simulated value of the $p$-th parameter in $\bm\phi$, respectively. For this covariance function, $\sigma^2$ is the GP marginal variance and $\delta_p$ defines the smoothness of the relationship between two values that are ``close together'' in dimension $p$. For high values of $\delta_p$ the correlation between the two conditional expectations with similar values for $\phi_p$ is small. The $\delta_p$ values are also treated as hyperparameters to be estimated from the data. 

Combining equations (\ref{NB_GP}) and (\ref{g_GP}), we can directly model the ``observed'' vector of net benefits as
\begin{equation}
\left(\begin{array}{c}
\mbox{NB}_t(\bm\theta_1)\\
\mbox{NB}_t(\bm\theta_2)\\
\vdots\\
\mbox{NB}_t(\bm\theta_S)
\end{array}\right) \sim \mbox{Normal}(\bm H \bm\beta, \bm{\mathcal{C}}_{\rm{Exp}}+ \sigma_\varepsilon^2\bm I). \label{NB_regr}
\end{equation} 
The model in (\ref{NB_regr}) includes $2P+3$ hyperparameters: the $P+1$ regression coefficients $\bm\beta$, the $P$ smoothness parameters $\bm\delta=(\delta_1,\ldots,\delta_{P})$, the marginal variance of the GP $\sigma^2$ and the residual error $\sigma_\varepsilon^2$, also known as ``nugget variance''. In this setting, therefore, the PSA samples for $\bm\phi$ are the ``covariates'' used to fit the non-parametric regression model, while the ``response'' is represented by the net benefits. Given that the estimation of the hyperparameters is the most expensive component of fitting a GP \cite{StrongOakley:2014}, in computational terms the efficiency of this method to estimate the EVPPI depends on the number of parameters of interest. 

Strong et al.\ \cite{StrongOakley:2014} integrate out $\bm\beta$ and $\sigma$ analytically and use numerical optimisation to calculate the posterior mode of the other hyperparameters $\bm\delta$ and $\sigma_\varepsilon$ analytically. This allows for great flexibility but at a computational cost in the order of $S^3$. As PSA is usually based on relatively large number of simulated values from the posterior distributions of $\bm\theta$ (\ie in the thousands) this procedure still takes considerable computational effort despite producing a significant improvement over MC methods, especially for larger numbers of parameter of interest. 

\section{Fast computation of the EVPPI using Integrated Nested Laplace Approximation}\label{our_stuff}
\subsection{Spatial statistics and Stochastic Partial Differential Equations}\label{SPDE}
An interesting application of GP regression is in the field of spatial statistics, where measurements are taken at different points in a spatial domain. For example, these can be the cases of influenza at locations in a geographical area (\eg a country) or the level of pollution at different monitoring sites. The main assumption in spatial statistics is that points that are ``closer'' to each other in a geographical sense share more common features and are influenced by common factors than those ``further away'' \cite{Tobler:1970}. 

A very popular specification of a spatial model when exact locations are available is based on the Mat\'ern family of covariance functions~\cite{Cressie:1993}, defined by
\begin{equation*}
\mathcal{C}_{\rm{M}}(r,s) = \frac{\sigma^2}{\Gamma(\nu)2^{\nu-1}}(\kappa \lVert \bm\phi_r-\bm\phi_s\rVert)^{\nu} \mbox{K}_\nu(\kappa \lVert  \bm\phi_r-\bm\phi_s\rVert), \label{Matern}
\end{equation*} 
where $\bm\xi=(\sigma,\kappa,\nu)$ is a vector of hyperparameters, $\lVert .\rVert$ denotes the Euclidean distance and K$_\nu$ is the modified Bessel function of the second kind and order $\nu$. The Mat\'ern covariance function is related to the covariance function in (\ref{ExpCov}), which can be obtained when $\delta_p$ is constant for all $p=1,\ldots,P$ and $\nu \to \infty$ \cite{Rasmussen:2006}.  This implies that the resulting covariance matrix for a specific set of input values is still dense, \ie a large number of the matrix elements are non-zero, which in turn generates a computational cost for matrix inversion in the order of $S^3$. 

However, Lindgren et al.\ \cite{Lindgren:2011} demonstrate that a sparse matrix can be used to approximate a GP with a Mat\'ern covariance function (Mat\'ern GP) by using Stochastic Partial Differential Equations (SPDE), which in general leads to a computation time for matrix inversion in the order of $S^{\frac{3}{2}}$. It can be shown that, in addition to being defined in terms of a relationship with the multivariate Gaussian, a Mat\'ern GP is also exactly equal to the function $g_t(\bm\phi)$ that solves a specific SPDE of the form

\begin{equation*}
\tau (\kappa^2 - \Delta)^{\frac{\alpha}{2}}  g_t(\bm\phi) = \mathcal{W}(\bm\phi),
\end{equation*} 
where $\mathcal{W}$ is Gaussian noise, $\Delta$ is the Laplacian operator, $\alpha = \nu +\frac{P}{2}$ (with $P=2$, in the spatial context) and the marginal variance of the Mat\'ern GP is 
\begin{equation*}
\sigma^2= \frac{\Gamma(\nu)}{\Gamma(\alpha) (4\pi)^{\frac{P}{2}}} \kappa^{-2\nu} \tau^{-2}.
\end{equation*}
Therefore, finding the solution to this SPDE is exactly equivalent to finding the function $g_t(\bm\phi)$, which as mentioned in \S \ref{GP} is instrumental in estimating the EVPPI. 

\begin{figure}[!h]
\centering
\includegraphics[width=8cm]{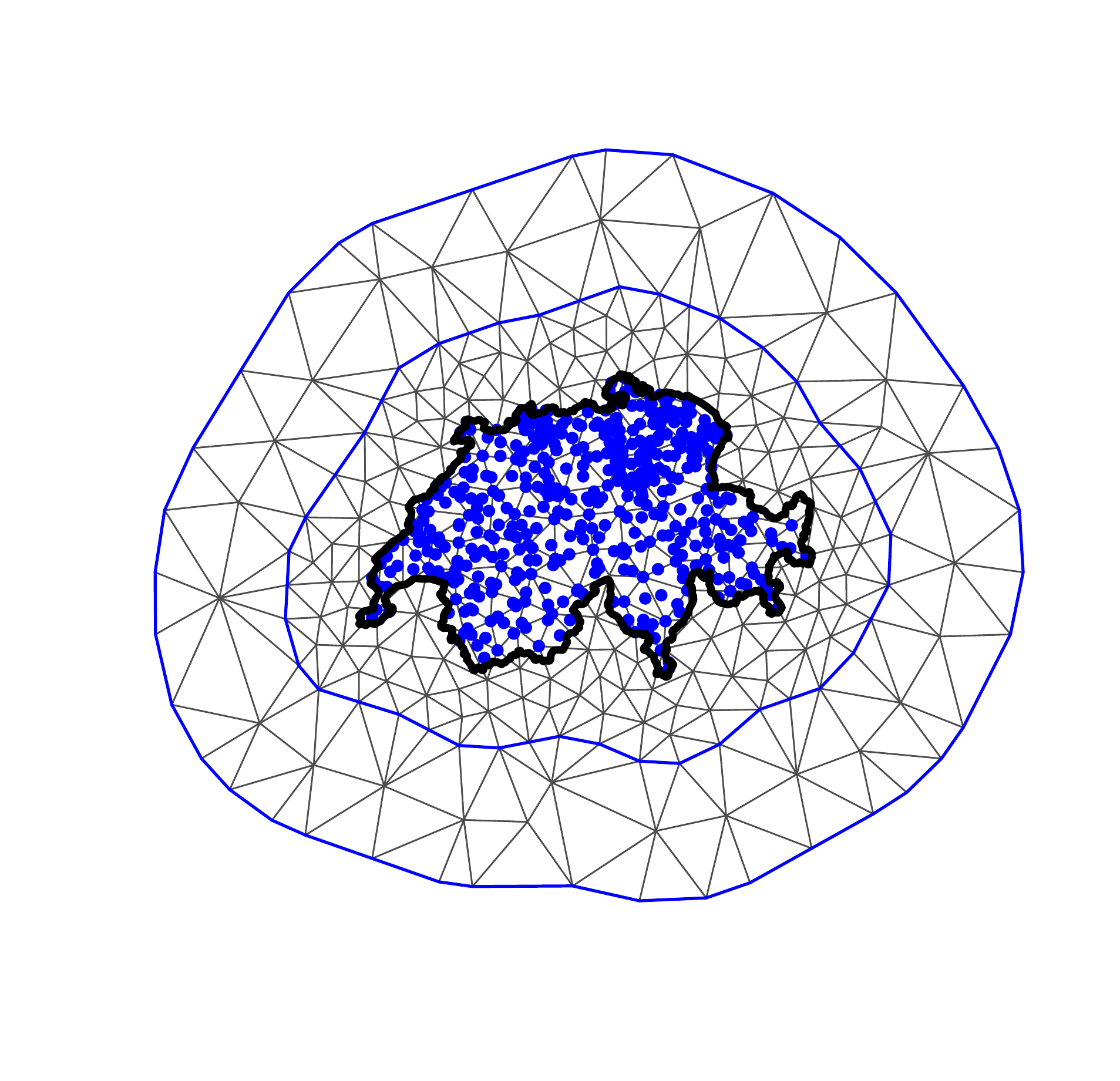}
\caption{An example of the grid approximation used to approximate the Mat\'ern GP in a proper spatial problem. The thick black line represent the border of Switzerland. The blue dots represent the positions where data points have been observed. These data points are used to estimate the value of the Mat\'ern GP throughout the geographical space (\ie the whole area covered by Switzerland, in this case).}
\label{Grid}
\end{figure}

The fundamental implication of this result is that efficient algorithms for solving SPDEs can be used to approximate the Mat\'ern GP. In practice, the SPDE is solved using the finite element method \cite{Ciarlet:1978}. First, the region over which the SPDE is being solved, \ie the range of $\bm\phi$, is split into small areas. In the proper spatial 2-dimensional case, a grid of small triangles is used; an example of this triangulation relating to the amount of rainfall in Switzerland over a pre-specified time horizon is shown in Figure \ref{Grid}. 
Notice that there are two triangulation boundaries with the border of Switzerland added for clarity. An inner boundary encases (or ``hugs'') all the data points relatively tightly, while the outer boundary is further from the points. This is because boundary conditions are imposed on the SPDE solver and it is important to avoid these impacting on the estimation of the smooth function $g_t(\cdot)$. The value of the Mat\'ern GP is then approximated by simple (linear) functions within each small triangular area. Therefore, the triangles are more tightly packed within the inner boundary to give a good approximation to the Mat\'ern GP, while in the outer region the grid can be rougher as the approximation is not as important.

The Mat\'ern GP is directly approximated at each intersection where triangles meet (\ie the vertices). Then, within each triangle it is approximated using linear interpolation by taking a weighted sum of the weights at the three corners of the triangle. Therefore, the vertex weights $\bm\omega$, which are estimated from the data (\ie the net benefit values, in this case), entirely determine the value of the Mat\'ern GP. Lindgren et al.\ \cite{Lindgren:2011} demonstrate that these vertex weights can be assumed to have a multivariate Gaussian distribution with a specific precision matrix, conditional on $\kappa$ and $\tau$. This precision matrix is, to a very good degree of approximation, sparse as non-zero entries correspond loosely only with points that ``share a triangle''.

\subsection{Computing the EVPPI using SPDE-INLA}\label{method}
Assuming a Mat\'ern covariance function, the model in (\ref{NB_regr}) becomes 
$\mbox{\textbf{NB}}_t \sim \mbox{Normal}(\bm H\bm\beta,\bm{\mathcal{C}}_{\rm{M}}+\sigma^2_\varepsilon\bm I)$, which can be equivalently re-expressed as
\begin{equation}
\mbox{\textbf{NB}}_t 
\sim \mbox{Normal}(\bm H\bm\beta +F(\bm\omega), \sigma^2_\varepsilon \bm I),\label{NB_INLA}
\end{equation} 
where $\bm\omega=\left(\omega_1,\ldots,\omega_S\right)'\sim \mbox{Normal}\left(\bm 0, \bm Q^{-1}(\tau,\kappa)\right)$ are the vertex weights and $\bm Q(\tau, \kappa)$ is the sparse precision matrix determined by the SPDE solution. The function $F(\cdot)$ maps from the position of the $\bm\omega$ values to the position of the data points on the grid. As is possible to see in Figure \ref{Grid}, many of the points do not lie exactly on the intersections between the triangles and are therefore calculated as a function of several $\bm\omega$ values.

Interestingly, the specification in (\ref{NB_INLA}) is in fact a \textit{Latent Gaussian Model} \cite{RueMartino:2009} as the response of interest $\mbox{\textbf{NB}}_t$ is dependent on a latent Gaussian vector $\bm\omega$ controlled by a small number of hyperparameters ($\kappa$ and $\tau$). This means that inference can be performed in a very efficient way by using the Integrated Nested Laplace Approximation (INLA) algorithm \cite{RueMartino:2009} (cf.~Apprendix \ref{INLA-appendix}), programmed in the \texttt{R} package \texttt{R-INLA} \cite{INLA}, which also includes extensions to implement the SPDE method \cite{Lindgren:2011,Lindgren:2012,Simpsonetal:2011,LindgrenRue:2013,Simpsonetal:2012,KrainskiRue:2013}. 

The SPDE-INLA method has been developed and successfully applied in a spatial context \cite{Simpsonetal:2012,Camelettietal:2011,Camelettietal:2013}, where inputs are proper coordinates (\ie longitude and latitude, hence defined in a 2-dimensional space) which makes the GP approximation extremely fast. However, calculating the EVPPI relies on a set of much higher dimensional inputs. While in theory the SPDE machinery works in higher dimensional spaces, the computational advantages will diminish in these cases. 

To fully exploit the computational savings of the SPDE-INLA procedure, we re-express the problem of computing the EVPPI in a ``spatial'' context. In this case, the simulated parameter vector for $\bm\phi$ designates a point in the $P$-dimensional parameter space. We consider that the net benefit, calculated as a function of $\bm\phi$, has been ``observed'' at this point. We then wish to find a representation of these $P$-dimensional points in at most 2-dimensional space, so that we can efficiently estimate the Mat\'ern covariance of this representation using the SPDE-INLA methodology. If no such projection exists then the computational complexity of fitting a Mat\'ern GP with a high dimensional SPDE means that the standard GP method is more appropriate.

As this projection will be used to predict the net benefit values, it makes sense to use a regression-based dimension reduction method. This class of methods tries to find a \emph{sufficient} reduction, \ie one for which the projection contains the relevant information about the net benefit function. Formally, we can express this condition as $\mbox{\textbf{NB}}_t \perp\!\!\!\perp \bm\phi \mid R(\bm\phi)$, where $R(\cdot)$ is the reduction function from $P$, the length of $\bm\phi$, to $d$, the number of dimensions needed to capture all the information in $\bm\phi$. There is a wealth of methods that can be used to estimate this sufficient reduction \cite{Li:1991,CookNi:2005,CookWeisburg:1991,LiWang:2007,Cook:2007,CookForzani:2008}. Specifically, we focus on Principal Fitted Components \cite{Cook:2007,CookForzani:2008} to calculate the EVPPI.

\subsubsection{Principal Fitted Components}\label{pfc}
Principal Fitted Components (PFC) is a model based inverse regression method. This means that in order to find a sufficient reduction we consider a model for $\bm\phi$ as a function of $\mbox{\textbf{NB}}_t$. As different models can be specified and lead to different sufficient reductions, the best fit model amongst a set of candidates should be chosen before finding the sufficient reduction.

The general form of these PFC models is based on a linear structure
\[\bm\phi = \bm\mu\ + \bm\Upsilon \bm f(\mbox{\textbf{NB}}_t) + \bm\epsilon,\]
where $\bm\mu$ is the intercept; $\bm\Upsilon$ is a $P \times d$ matrix to be estimated to determine the sufficient reduction; $\bm f(\cdot)$ is a vector-valued function of $\mbox{\textbf{NB}}_t$; and $\bm\epsilon$ is an error term in the exponential family. The form of the error structure changes the way in which the reduction is calculated and methods have been developed for independent, heteroskedastic and unstructured errors. 

In order to use PFC, the function $\bm f(\cdot)$ and the error structure need to be specified. In our problem, we consider normally distributed errors and set $\bm f\left(\mbox{\textbf{NB}}_t\right) = \left[\alpha_1\mbox{\textbf{NB}}_t,\alpha_2\mbox{\textbf{NB}}_t^2,\dots,\alpha_h\mbox{\textbf{NB}}_t^h\right]^T$, although in general $\bm{f}(\cdot)$ can map to any function of $\mbox{\textbf{NB}}_t$. Additionally, the number of dimensions $d$ needed to capture all the relevant information in $\bm\phi$ needs to be specified. It is then advisable to select the values of $d$ and $h$ (the polynomial degree) associated with the best performing inverse regression specification, \eg in terms of model fitting, as measured by information criteria such as the AIC \cite{CookForzani:2008}. 

The cost of fitting any of these models individually is negligible and thus fitting a number of models in correspondence to a set of chosen $d$ and $h$ values adds very little to the computational time required to estimate the EVPPI. In any case, because PFC assumes that the number of dimensions needed to capture the information in $\bm\phi$ is no larger that then number of dimensions in the function $\bm{f}(\cdot)$, for simple relationships between the net benefit and $\bm\phi$, the sufficient regression is low~dimensional. 

To fully exploit the spatial nature of the SPDE-INLA methodology, $d$ must be set to 2. However, identifying the optimal value for $d$ allows us to check the validity of the EVPPI approximation. If one dimension is sufficient to capture all the information in $\bm\phi$, there is no harm in using a second component because this will only add information and the reduction will remain sufficient. On the other hand, using two dimensions when the AIC suggests $d>2$, may lead to a loss in information. Consequently, the EVPPI estimate based on a two-dimensional reduction of $\bm\phi$ may be biased. In light of the large computational savings and the fact that the AIC has a tendency to overestimate $d$ \cite{CookForzani:2008}, it may still be worth using the projection to estimate the EVPPI and then perform thorough model checking (\eg by means of the residual plots) to assess its performance, before resorting to more computationally intensive methods. We return to this point in \S\ref{results}.

From the theoretical point of view, PFC provides a robust method for determining the sufficient reduction \cite{CookForzani:2008}. Thus, combining PFC and SPDE-INLA to estimate the EVPPI seems to be a valid strategy. Additionally, due to the flexibility of the INLA algorithm it is possible to cater for more complicated structures in the relationships between the net benefit and the parameters of interest, as we discuss in \S\ref{tech}.

We have combined PFC and SPDE-INLA, along with some simple PFC model selection, in the \texttt{R} package \texttt{BCEA} \cite{BCEA:2013,BaioBCEA:2012} to allow users to integrate standard economic analysis with efficient calculations for the EVPPI in large dimensions. This function relies on the \texttt{R-INLA} and \texttt{ldr} packages \cite{INLA,ldr}. We have also implemented model checking procedures for the non-parametric regression as standard in order to aid practitioners. This potentially improves the use of value of information analysis as a tool for PSA in applied health economic problems. 


\section{Examples}\label{results}
We present two case studies of health economic models and compare the estimates of the EVPPI using the direct GP regression implemented by Strong \textit{et al.} and our SPDE-INLA projection method. For both case studies, random subsets of between 5 and 16 parameters of interest were considered to compare the performance of the GP procedures --- notice that this represents the standard range of parameter subsets that would be used practically for EVPPI calculation using GP \cite{StrongOakley:2014,Dongetal:2007}. For each subset, the EVPPI was calculated using both methods and for a willingness-to-pay threshold of $k=20000$ monetary units, say~\pounds. The computational time and EVPPI estimate was then recorded for both methods to allow a direct comparison. 

\subsection{Vaccine Study}
The first case study (referred to as the ``Vaccine study'') is a Bayesian health-economic model proposed to analyse the effect of an influenza vaccine on health outcomes and costs. A more detailed description of the example is presented in \cite{Baio:2012}. The parameters are sampled from their joint posterior distribution using Markov Chain Monte Carlo (MCMC) methods. These sampled parameter values are used to calculate the net benefits and~the~EVPPI.

Two treatment options are considered, either the vaccine is available to the population ($t=1$) or not ($t=0$). If an individual gets influenza, they are treated with anti-viral drugs and will often visit the doctor. Complications may occur, including pneumonia and hospitalisation, in which case there will be some indirect costs such as time off work. The cost of the treatment is the acquisition cost of the drugs, the time in hospital, the doctor's visits and the cost of the vaccine. The benefit of the treatment is measured in QALYs, to which each adverse effect contributes negatively.

The Vaccine model includes 28 key parameters representing the probability of infection, the reduction in risk due to the vaccine, the occurrence of complications, the monetary costs of the interventions and the QALY loss due to different health states. However, as the model is built as an evidence synthesis, additional sources of uncertainty are present; for instance, the true number of people getting influenza or the true number of people getting side effects are unknown. Considering all the unobserved quantities in the model, the number of parameters increases to 62.

\subsection{SAVI Study}
The second case study is a simple fictional decision tree model with correlated parameters, presented at the Sheffield Accelerated Value of Information web app \cite{SAVI:2014} (hence this example is referred to as the ``SAVI study''). The model has two treatment options and 19 underlying parameters. A more in-depth model description is presented in \cite{Brennanetal:2007}. Most importantly, the 19 underlying parameters are assumed to follow a multivariate Gaussian distribution and thus the conditional distribution $p(\bm\psi \mid \bm\phi)$ can be computed analytically and the EVPPI can therefore be calculated using MC simulation. The SAVI web app provides 10\,000 PSA samples of all the 19 parameters, along with the simulated costs and benefits of both treatment options. The number of available PSA samples poses a significant challenge for the standard GP regression method. For this reason, only the first 1\,000 observations are used for the comparison with our~SPDE-INLA~method. 

\subsection{Computational Time}\label{comptime}
We begin our discussion of the two EVPPI estimation methods by comparing the computational time required to obtain an estimate. The EVPPI estimates were calculated using 1\,000 PSA samples for both case studies and both methods. To compare our SPDE-INLA method we used the code available from Strong \cite{Strong:2012:Code} with a slight modification. This modification changed the numerical optimiser (used to estimate the hyperparameters) to give quicker computation time and more accurate results although in some cases this optimiser can struggle numerically and the slower optimiser must be used.

Additionally, to allow for a fair comparison between the two methods only 500 PSA runs were used to estimate the hyperparameters by numerical optimisation. This is because for each  step, an $S \times S$ dense matrix must be inverted. As this is computationally expensive, it is suggested \cite{StrongOakley:2014} that the full PSA run is not used to calculate the hyperparameters. Once the hyperparameters have been estimated, all 1\,000 PSA samples are used to find the fitted values $\hat{g_t}(\bm\phi_s)$, so all the information is utilised. Using all 1\,000 observations for the optimisation step can give more accurate results and is sometimes necessary (see for example \S\ref{accuracy}). 

\begin{table}[!h]
\caption{The computational time required (in seconds) to calculate an EVPPI using both the GP regression method and SPDE-INLA method for increasing numbers of parameters for both case studies\label{time}}
\centering
\begin{tabular}{ccccc}
\hline
\bf{Number of parameters } & \multicolumn{4}{c}{\bf{Computation time (seconds)}} \\
\textbf{of interest} & \multicolumn{2}{c}{\bf{Vaccine Example}} &\multicolumn{2}{c}{\bf{SAVI Example}}\\
\hline
&GP &SPDE-INLA&GP&SPDE-INLA \\
\hline
5&17&9&17 &7 \\
6&42&10&14 & 7\\
7&45&10&18 &7\\
8&57&11&21 & 8\\
9&74&8&26 & 8\\
10&86&8&31 &9\\
11&70&7&37& 8\\
12&60&8&47 &8\\
13&84&11&52&7\\
14&188&8&66 & 6\\
15&470&7&70 &7\\
16&121&8&71 & 7\\
\hline
\end{tabular}
\end{table}

The computational time for the GP regression increases substantially with the number of parameters of interest, between 17 and 470 seconds for the Vaccine case study and 17 and 71 seconds for the SAVI example. However, interestingly, the computation time does not increase uniformly for GP regression. This is due to the numerical optimisation, as occasionally additional steps are required to reach convergence.

The computation time for our SPDE-INLA method remains constant as the number of parameters increases. The computation time of our SPDE-INLA method is significantly lower than the GP regression method, up to around 70 times faster. Even for 5 parameters of interest, it is between 2 and 2.5 times faster, despite the fact that we are using all the data points to estimate the EVPPI, albeit from a projected input space.

To understand if our method scales to larger PSA datasets, EVPPI estimates using all 10\,000 PSA samples from the SAVI case study were also calculated. The computational time required to calculate an EVPPI estimate was between 40 and 80 seconds with an average time of 56 seconds. This is significant as the computation time does not increase exponentially using the SPDE-INLA method; the computation time is between than 6 and 10 times slower for a 10 fold increase in number of PSA samples. Crucially, the speed of our SPDE-INLA method depends on the density of the grid approximation. Therefore, its computational effort could be decreased by using a sparser grid, although this would clearly affect the quality of the EVPPI estimate. It would, therefore, be possible to use our method to calculate the EVPPI for larger PSA data sets. This may be relevant, for instance, in models involving individual level simulations (often referred to as ``microsimulations'' in the health economic literature) \cite{Weinstein:2006}, where larger PSA samples are required to fully assess the underlying distributions.

\subsection{Accuracy}\label{accuracy}
In general, it is difficult to establish whether an estimate of the EVPPI is accurate since the calculations of the EVPPI are frequently analytically intractable. In fact, for the Vaccine model there is no closed-form expression for the EVPPI, while, given its simplified model structure, for the SAVI example long MC runs can establish the EVPPI up to an inconsequently small Monte Carlo error. Thus, it is difficult to determine which method is more accurate when the two approximate EVPPI values diverge, as no baseline comparator is easily available. Nevertheless, there are at least two potential features that we can use to assess the reliability of our estimates. 

\subsubsection{Monotonicity with respect to the number of parameters of interest}
It can be easily shown that the EVPPI is a non-decreasing function of the number of parameters of interest (cfr.\ a proof in appendix \ref{monotone}). This means that, provided the smaller subsets are entirely contained within the larger subsets, the EVPPI estimates should be non-decreasing. This property provides a possible assessment of the accuracy of the methods: if one method fulfils this property and the other does not, then the former could be more accurate. It is important to note that monotonicity is only a necessary condition for a good EVPPI estimate and not a sufficient one. It is clearly possibly to construct a function that gives a monotone sequence, such as simply giving the number of parameters in the set, but clearly does not estimate the EVPPI at all.

\begin{figure}[!h]
\centering
\subfloat[]{\includegraphics[scale=.55]{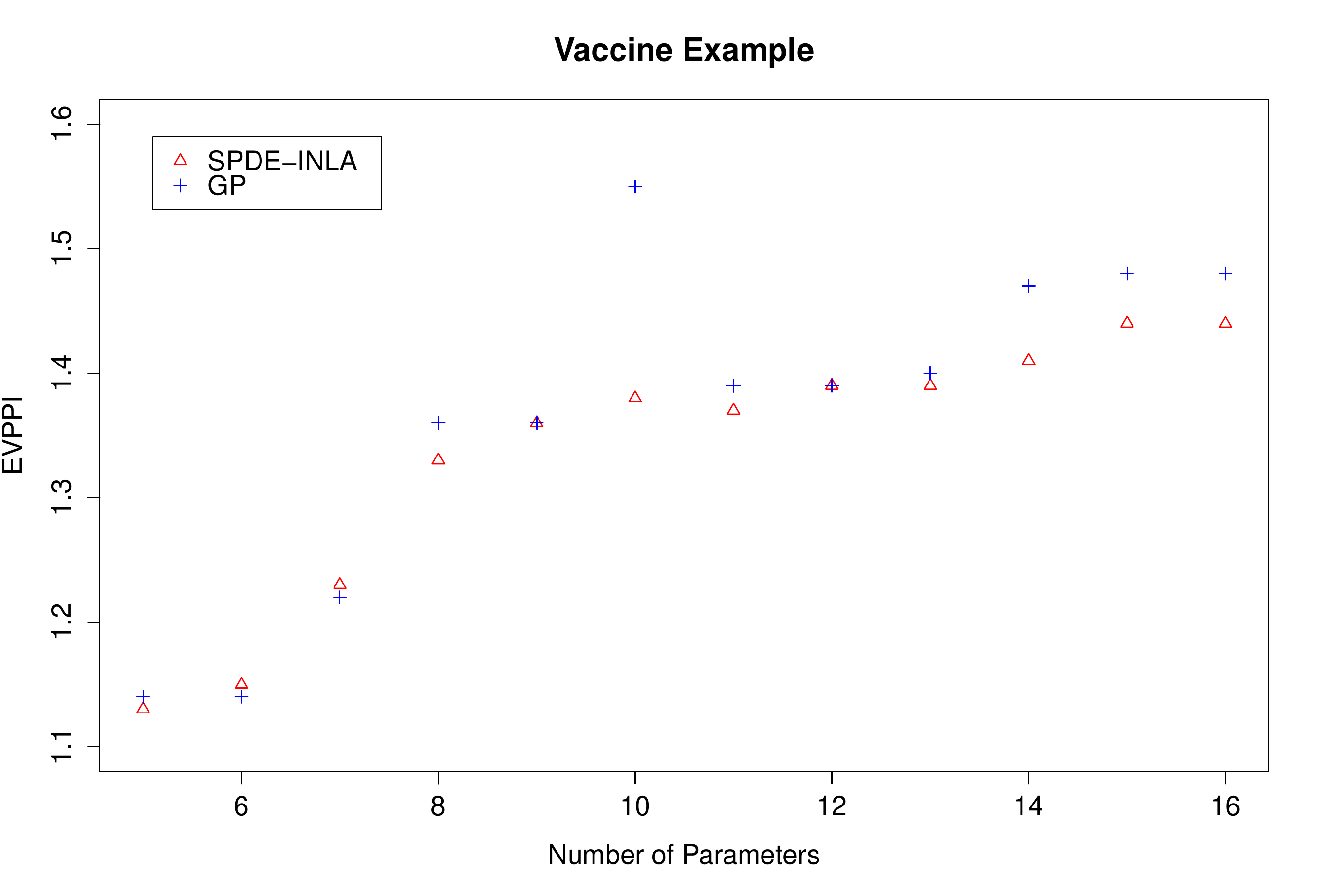}} \hfill
\subfloat[]{\includegraphics[scale=.55]{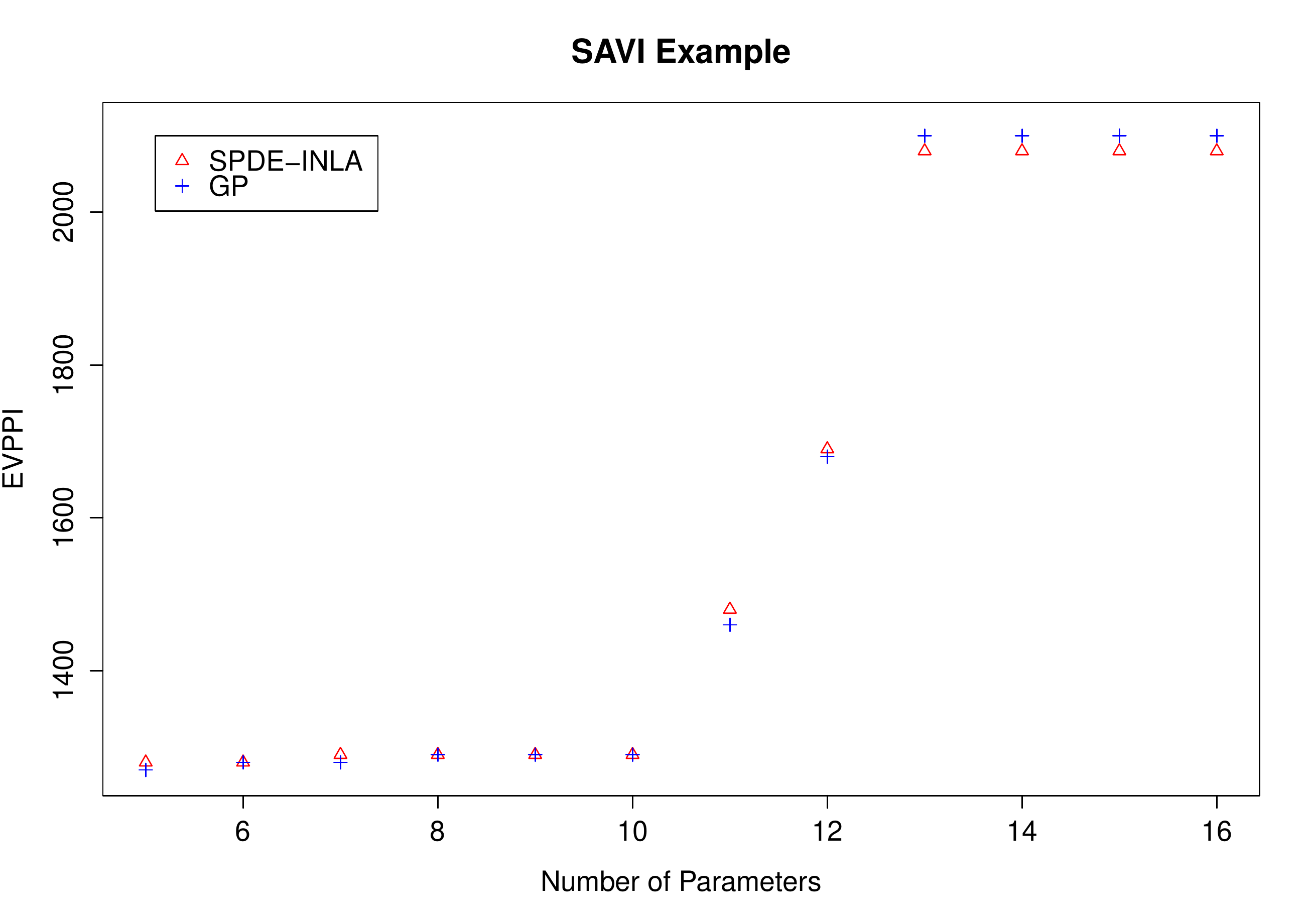}}
\caption{The EVPPI estimate for the Gaussian Process regression method (GP) and the new method developed in this paper (SPDE) for increasing parameter subset size for the Vaccine (panel a) and the SAVI (panel b) case studies.}\label{table}
\end{figure}

Figure \ref{table} shows the EVPPI estimate for increasing parameter subset sizes for both case studies. The smaller sets of parameters of interest are simply subsets of the larger sets of parameters.
For the Vaccine example, shown in panel (a), the standard GP regression method has some difficulty retrieving monotonicity, specifically for the parameter set containing 10 parameters which is clearly overestimated.  Our SPDE-INLA method also overestimates the EVPPI for the set containing 10 parameters, but by 0.01 or less than 1\%. This given an indication of the accuracy of our EVPPI estimation method. This over-estimation for the standard GP method is due in part to the incorrect estimation of the hyperparameters based on the reduced PSA dataset. If the estimate is obtained using all 1\,000 PSA samples then the EVPPI is 1.39, which respects monotonicity but the computation time increases to 256 seconds compared to 86.

For the SAVI example (panel b) the monotonicity is respected across both methods and the EVPPI values, rounded to 3 significant figures, are similar. For both examples, the EVPPI values for the smaller parameter subsets are similar across both methods. As the length of $\bm\phi$ increases the SPDE-INLA method underestimates slightly compared to the standard GP but only by at most 3\% of the total EVPPI value. There is evidence that the SPDE-INLA method is accurate while possibly guarding against spurious results that come from estimating the hyperparameters based on a smaller subset of the PSA samples. We note however that some issues with underestimation may occur with larger subsets.

\subsubsection{SPDE-INLA as EVPI approximation}
To investigate whether the SPDE-INLA method underestimates the EVPPI for larger parameter subsets, we compare the results from our method to the overall EVPI, which represents the largest parameter subset available for each example. As mentioned earlier, the overall EVPI \eqref{EVPI-eq} can be easily calculated directly from the available PSA data, since as there are no nuisance parameters we can use a single loop to estimate it. We can then use our method to calculate the overall EVPI, by considering that all the underlying model parameters are of interest. This allows us to compare our method directly with the ``true''  MC EVPI estimate. The EVPI was calculated for both cases studies and the results are shown in Table \ref{EVPI}. The computational time required to calculate these estimates is 14 seconds for the Vaccine example and 8 seconds for the SAVI.

\begin{table}[!h]
\caption{The EVPI values calculated using the PSA samples directly and our SPDE-INLA method\label{EVPI}}
\centering
\begin{tabular}{ccc}
\hline
\bf{Case Study} & \bf{MC EVPI} & \bf{SPDE-INLA EVPI}\\
\hline
Vaccine &2.52 &2.51\\
SAVI&  2100 & 2080 \\
\hline
\end{tabular}
\end{table}

For both case studies, the SPDE-INLA approximation is correct to two significant figures, with a small discrepancy in the third significant figure. This gives a further indication that the EVPPI estimated using our method is an accurate method for calculating the EVPPI, possibly demonstrating that even for large numbers of parameters of interest the underestimation of the EVPPI is not severe. 

For the Vaccine example, there are 62 parameters that contribute to the model uncertainty. We are therefore approximating the Mat\'ern covariance function with a projection from a 62- to a 2-dimensional space. However, despite the difficulty of preserving the original data structure with this projection, the AIC suggests that this is a sufficient reduction and the EVPI estimate is still very close to the true value. 

\subsubsection{Analytic Results}
Since the parameters for the SAVI case study are in fact drawn from a known multivariate Normal distribution, the EVPPI can be calculated using a single MC loop. The parameters of interest, $\bm\phi$, are sampled from their posterior marginal and then the conditional expectation for each simulated parameter vector is calculated analytically. Strong et al.\ \cite{StrongOakley:2014} provide these EVPPI values, based on a long MC run, for three different parameters subsets of size 2 and 4, which can be used to test the accuracy of our procedure. 

Both GP regression and GAM can be used to calculate the EVPPI using all 10\,000 avalible PSA samples and the willingness-to-pay is fixed at \pounds 10\,000. Table~\ref{true} gives the single MC loop EVPPI together with all three estimated values, GAM, GP regression and SPDE-INLA. It is clear from this table that our method's performance is in line with the other approximation methods while all 3 methods overestimate the true EVPPI value in this setting.

\begin{table}[!h]
\caption{Comparison of the EVPPI estimation methods, standard GP, GAM regression and the SPDE-INLA method with ``true'' EVPPI values based on $10^7$ Monte Carlo simulations.}
\label{true}
\centering
\begin{tabular}{ccccc}
\hline
&\multicolumn{4}{c}{\bf{EVPPI estimate (Time to compute in seconds)}}\\
\bf{Parameter Subset}&\bf{MC Simulations}&\bf{GP}&\bf{GAM}&\bf{SPDE-INLA}\\
\hline
2 Parameters - $\phi_5,\phi_{14}$&248 &274 (375)&277 (0.78)&278 (41)\\
4 Parameters - $\phi_5,\phi_6,\phi_{14},\phi_{15}$ & 841 & 861 (367)&862 (98)&856 (48)\\
2 Parameters - $\phi_7,\phi_{16}$ &536&549 (390)&546 (0.25) &549 (43)\\
\hline
\end{tabular}
\end{table}

The computational time required to calculate these estimates are given in Table \ref{true}. Clearly, for the 2 parameter setting the GAM regression method is the most appropriate as it takes under a second to calculate the EVPPI estimate. However, for the 4 parameter example, the computational time is lowest for our SPDE-INLA method, which takes half the computational effort of the GAM regression method and around 10\% of the standard GP. This demonstrates the computational effort required for the standard GP method using a large number of PSA samples, despite using only 500 PSA samples to find the hyperparameters (see \S\ref{comptime}).

\section{Technical Considerations}\label{tech}
There are several technical aspects relating to the implementation of this method that can affect its estimation performance. These considerations are mostly due to ensuring that the approximations used to estimate the EVPPI are not too rough, whilst retaining the computational advantages of the method. 

\begin{figure}[!h]
\centering
\subfloat[]{\includegraphics[width=8cm]{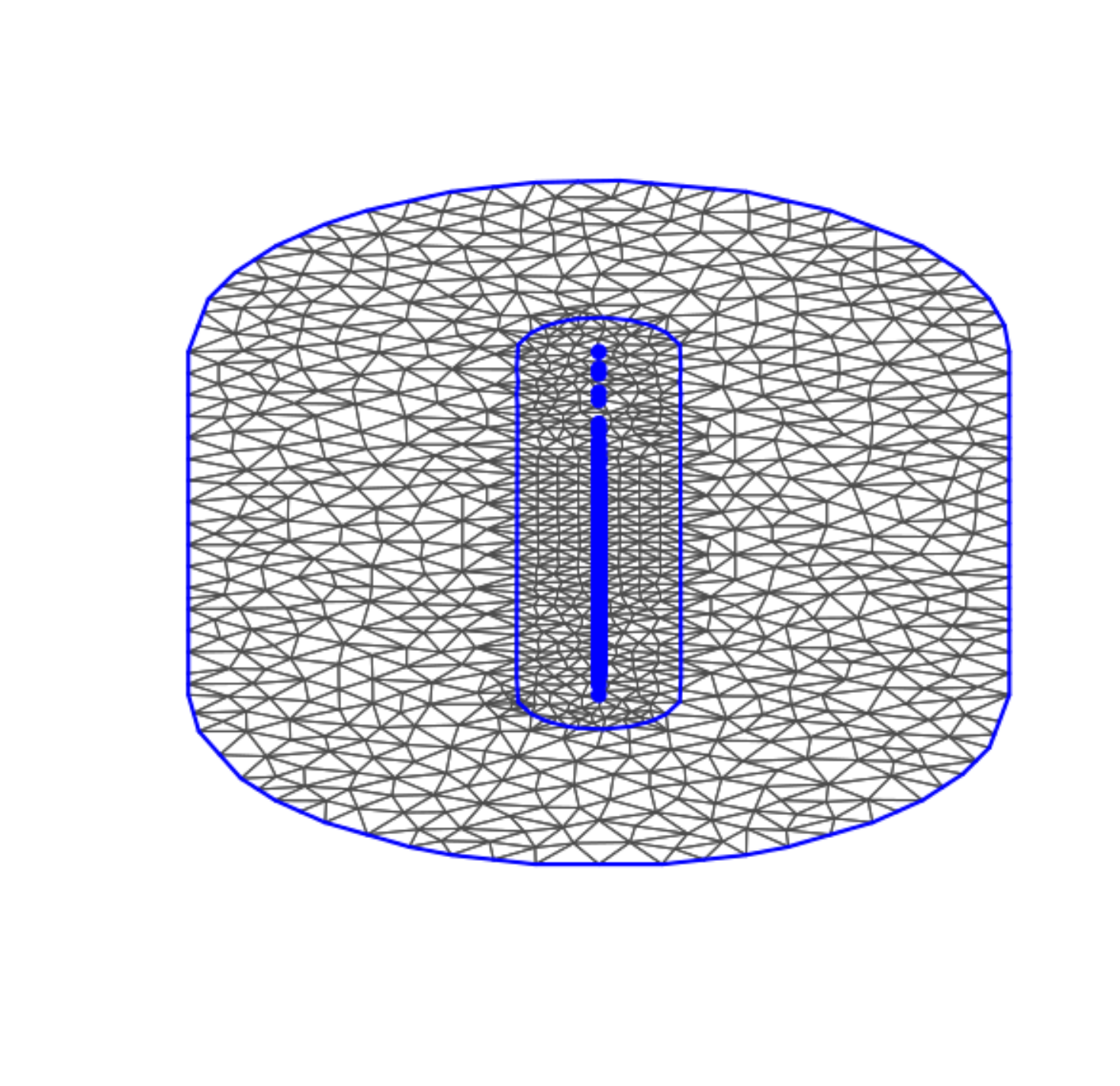}}
\subfloat[]{\includegraphics[width=8cm]{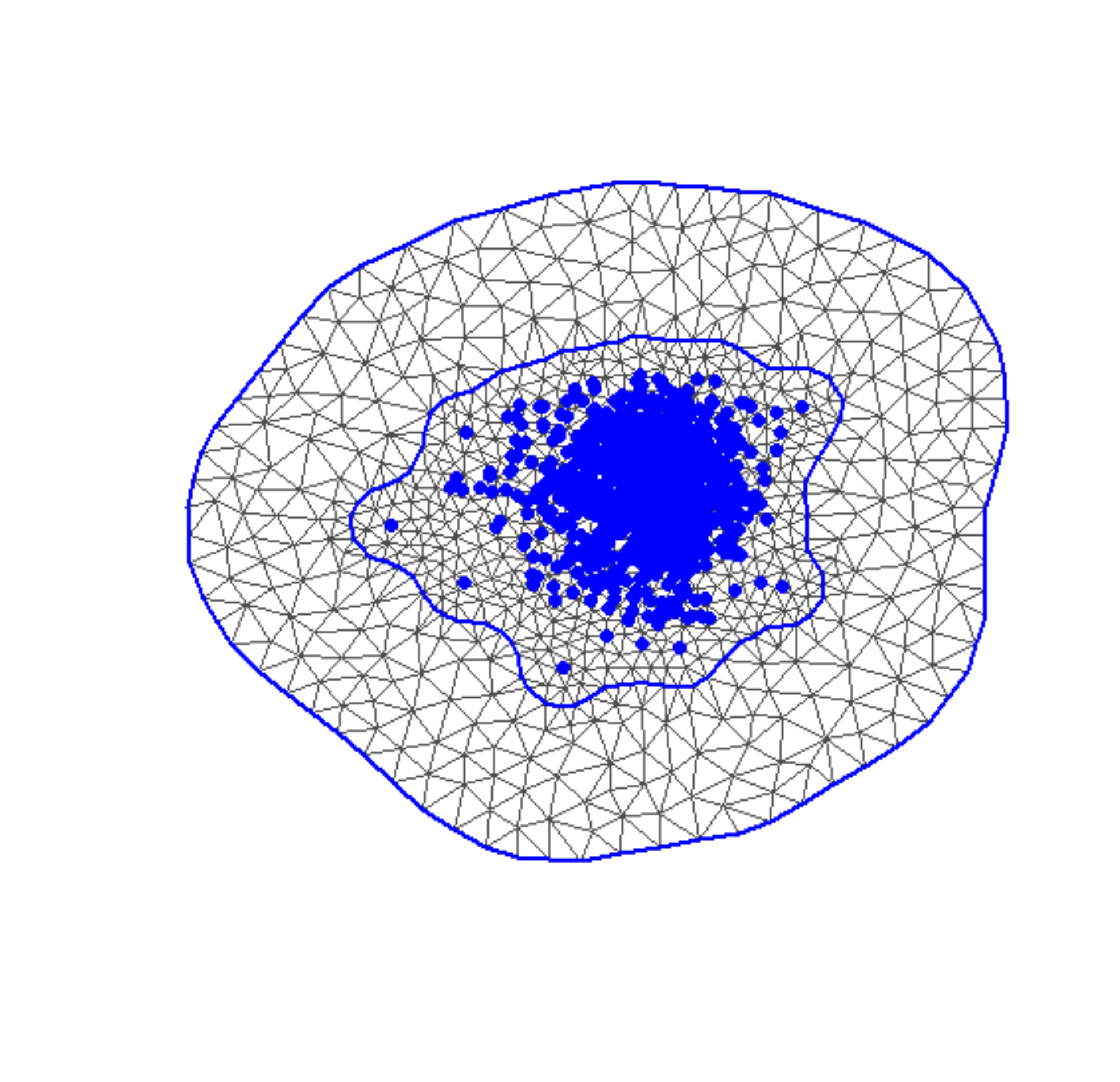}}
\caption{Two grid approximations for the same data set. The LHS shows the triangulation when the variables are left on their original scale, with the projected data points in blue. Notice that there are a large number of triangles in this case, but a relatively small number that surround the data points. In contrast to this, on the right, where the data points are scaled we note that a much larger number of mesh points cover the data, allowing for a more accurate Mat\'ern field approximation for a fixed computational time.} 
\label{wrong}
\end{figure}

The most important technical aspect of the SPDE-INLA procedure is the grid approximation used to build up the Finite Element Approximation to the Mat\'ern field. To create an accurate approximation, the triangulation must completely surround the data points with a significant distance from the outermost data point to the final boundary, as there are artificial constraints at the boundary of the triangulation. To reduce the computation time, a tight boundary hugs the data points closely and within this boundary the mesh points are dense to give a good approximation. Outside of this inner boundary, the approximation can be rougher and the triangles are therefore larger. The mesh approximation is most efficient when the two dimensions (coming from the projections for EVPPI calculation) are on approximately the same scale. Therefore, the PSA inputs should be rescaled before calculating the projection. This avoids situations such as that shown in Figure \ref{wrong} (a), where a large number of triangles cover an area with no observations. Rescaling has no effect on the estimated EVPPI value \cite{StrongOakley:2014} but does significantly decrease that computation time.

The triangulation must be dense enough to adequately capture the underlying structure or the EVPPI estimate will be incorrect. However, as the computation time of the method is directly related to the number of vertices, there is a trade-off between accuracy and computational time. Most importantly, larger datasets require denser triangulations to calculate the EVPPI efficiently and typically the number of mesh points should be greater than the number of data points. If some estimation difficulties are observed by a standard model checking procedure, such as residual plots, the estimation can sometimes be improved by making the triangulation more dense.

The inner boundary should completely encase the data points. However, sometimes extreme outliers can be isolated by this boundary and this affects the Mat\'ern field approximation. Therefore, care must be taken to ensure that all points lie within the inner boundary. However, as our ``data'', $\bm\phi$, are typically from PSA samples (such as MCMC samples), a relatively large number of observations are normally available and thus outliers are rare. An extreme outlier should probably be investigated thoroughly.

In some more complex examples, both estimation methods can struggle and therefore the small loss in flexibility from our method can slightly worsen its estimation performance. However, this drawback can be overcome by adding additional structure to the linear predictor. This means that the $\bm H$ matrix in Equation (\ref{H}) changes its form to include non-linear functions of the parameters. Mostly importantly, allowing for 2\textsuperscript{nd} or 3\textsuperscript{rd} order interactions between the parameters seems to account for the flexibility lost by using the projections whist retaining the computational advantages. Adding these extra terms fits easily into the INLA framework and so this can been easily added to the standard EVPPI calculation~method.

Finally, we reiterate that model checking should be thoroughly performed to ascertain when this additional flexibility should be used. Specifically, highly structured residuals indicate that the regression curve has not picked up all the relevant relationships present in the data \cite{Strongetal:2015,StrongOakley:2014}. Therefore, we advocate integrating model checking as a standard part of EVPPI estimation and stress that poorly fitted models result from either a lack of flexibility in the linear predictor or an incorrect PFC model.

Our implementation of the INLA-SPDE method to compute the EVPPI in the \texttt{R} package \texttt{BCEA} accounts for all these potential issues, including allowing the user to customise the linear predictor, and thus provides a reliable tool for practitioners. Options are also available to fine tune the model used to find the PFC and to use standard residual plots to check the fit of the non-parametric regression model \cite{BCEA:2013}.

\section{Conclusion}\label{discussion}
This paper develops a fast method for Gaussian Process regression in order to reduce the computational effort required to calculate the EVPPI for health economic evaluations. This method is based on a spatial interpretation of GP regression and projections into 2-dimensional space. This in turn allows us to use a fast computation method developed in spatial statistics, based on calculating a sparse precision matrix that approximates a Mat\'ern GP. Finally, this sparse precision matrix allows us to use the INLA methodology for fast Bayesian computation of the hyperparameters for the GP. It also allows us to find fitted values at no additional cost, which are then in turn used to estimate the EVPPI.

Despite the methodological complexity of our GP regression method, the user-friendly \verb"R" package \verb"R-INLA" can be used to estimate the hyperparameters and find the  fitted values. This simplifies the implementation of our method, allowing us to integrate it into a straightforward \texttt{R} function. This GP regression method significantly decreases the computation time required to calculate the EVPPI for larger subsets of  parameters of interest as it is at least 2 times faster than standard GP regression method, taking around 10 seconds to calculate an EVPPI estimate with 1\,000 PSA samples. 

There is little loss of accuracy when using our method in the examples we have considered. For larger subsets the EVPPI estimate is slightly underestimated compared to the standard GP methods, but this does not seem to be severe as the EVPI estimates are very close. Additionally, in some examples, our method seemed to be more accurate than the standard GP regression method, due to the breakdown of numerical~optimisation. These results are conditional on the dimension reduction being sufficient to capture all the information in $\bm\phi$. 

There are several important points of further research. Firstly, further work is required to understand the impact of our new method on the bias and standard error of the EVPPI estimate, especially considering the error introduced by using projections. These properties are important and estimation methods have been provided for the standard GP regression method \cite{StrongOakley:2014}. Secondly, a sparser grid could decrease the computational time required for this method. However, a comprehensive understanding of the impact of the density of the grid on the EVPPI estimate is needed. Ideally, we would determine an optimal grid density in terms of required computation time and accuracy. Finally, it is important to investigate how successful our method is compared to other fast GP regression methods. Lindgren et al.\ \cite{Lindgren:2011} demonstrate that the SPDE framework can be extended to non-stationary fields and thus, this method may provide quick GP regression for non-stationary processes.

Our method has the potential to have an important impact on the practice of health economic evaluation; the analysis of the value of information is well known as a potentially effective tool to determine research priority and the accuracy of decisions made as a result of economic models. Nevertheless, their practical applications has been thwarted by the complexity of the resulting calculations. Our method substantially reduces the computational time and is implemented in an \texttt{R} package, with stand-alone code also available from the authors, which means that practitioners and regulators can use it routinely to assess the impact of uncertainty in models on the decision-making process being investigated.

\subsection*{Acknowledgements}
Anna Heath is funded by the EPSRC. Dr Gianluca Baio is partially funded by a research grant sponsored by Mapi. Additionally, the authors would like to thank Alice Davis for helpful discussion on the SPDE-INLA methodology and two anonymous reviewers for their helpful comments. 



\bibliographystyle{wileyj}
\bibliography{bib}

\appendix
\section{The Integrated Nested Laplace Approximation}\label{INLA-appendix}
The Integrated Nested Laplace Approximation (INLA) is a fast approximate Bayesian inference method for a wide class of models known as Latent Gaussian Models (LGMs) \cite{RueMartino:2009}. This class of models is broad as many standard modelling scenarios can be reformulated as LGMs including regression models, dynamic models and spatio-temporal models~\cite{RueMartino:2009}.

A LGM is characterised by the fact that the data, $y_i$ can be defined by a parametric family with a parameter $\mu_i$ linked to a structured linear predictor $\eta_i$, based on a set of covariates $\bm\gamma_i=(\gamma_{i1},\ldots,\gamma_{iJ},\ldots,\gamma_{i(J+K)})$, through some link function~$h(\cdot)$: \begin{equation}h(\mu_i) = \eta_i = \alpha + \sum_{j=1}^{J} f_j(\gamma_{ij}) + \sum_{k=1}^{K} \beta_k \gamma_{ik} +\epsilon_i,\end{equation} where $f_j(\cdot)$ are unknown functions of the covariates $\bm\gamma_i$, $\bm\beta=(\beta_1,\ldots,\beta_K)$ are fixed regression coefficients, $\epsilon_i$ is some error term, and $J$ and $K$ are the number of functions of covariates and regressed covariates in the model \cite{Martins:2013}. The functions $f_j$ can be of any form and typically can represent autoregressive models, spatial effects or seasonal effects. In these settings the covariates $\bm\gamma_{i}$ give sequential or spatial information about the data $y_i$. A standard generalised linear model also fits this framework where all the functions $f_j(\cdot)$ are equal to $0$.

To complete the LGM formulation, a Gaussian prior is assigned to the set of parameters defining the linear predictor $\bm\theta=\left(\alpha,\bm\beta, f_j(\cdot),\epsilon_i\right)$, depending on some hyperparameters $\bm\lambda$ (typically, these determine the precision matrix of $\bm\theta$). Clearly, the number of elements in $\bm\theta$ is likely to be large and therefore, to allow for fast computation, INLA is restricted to the case where the Gaussian prior used has a ``sparse'' precision matrix. This Gaussian prior with a sparse precision matrix is also known as a Gaussian Markov Random Field (GMRF) \cite{RueHeld:2005}. Fast computation using INLA is ensured if $\bm\lambda$ contains a relatively small number of elements, typically no greater than 6.

At first glance, enforcing sparsity in the prior for $\bm\theta$ may seem restrictive as sparsity in the covariance matrix implies \emph{marginal} independence. However, sparsity in a precision matrix only enforces \emph{conditional} independence, a much looser restriction. A $0$ entry in the precision matrix implies that the two elements are independent conditionally on all other elements. The \emph{Markov} property encoded in this sparse matrix implies that the field is memoryless: values only depend directly on a few neighbours. In the SPDE-INLA setting these neighbours are those $\omega$ values which share a triangle.

Operationally, INLA explores the approximate joint posterior of the hyperparameters $\bm\lambda$ by determining the density of the Laplace approximation at a grid of points in the support of $\bm\lambda$. This grid is found by ``stepping'' along each axis of the hyperparameter space until the density falls below a specified threshold. The density of the Laplace approximation is then evaluated at each combination of these axis points; if the density at these points is above the threshold then the point is included in the grid. Interpolation is then used to approximate the posterior at all points in $\bm\lambda$. The posterior marginals for $\bm\lambda$ can then be found by using these lattice points for numerical integration.

The marginals for the parameters $\bm\theta$ are then approximated by another (simplified) Laplace approximation. This Laplace approximation is evaluated at each of the hyperparameter values on the lattice and the approximate marginals for $\bm\theta$ are given as a weighted sum of the Laplace approximation for each configuration of the hyperparameter set (weighted by the density at that point). In this sense, the approximate marginals for $\bm\theta$ are \emph{nested} within the Laplace approximation for posterior distribution of the hyperparameters. 

\section{Monotonic EVPPI estimates}
\label{monotone}
It can be easily demonstrated that the EVPPI is a non-decreasing function of the size of the parameter subset, provided the smaller subset is entirely contained within the larger subset. Firstly, some notation must be set up. In line with the paper, $\bm\theta$ represents the set of all underlying model parameters, $\bm\phi$ is the full set of  parameters of interest and $\bm\psi$ is the complement set, $\bm\theta=(\bm\psi,\bm\phi)$. In addition to this notation, define $\bm\xi \subset \bm\phi$ as a smaller subset of parameters of interest and $\bm\xi^c$ as the complement of this set such that $\bm\phi=(\bm\xi,\bm\xi^c)$. 

Using this notation we demonstrate that \[\mbox{EVPPI}(\bm\phi) \geq \mbox{EVPPI}(\bm\xi),\] where EVPPI$(\bm\phi)$ is the EVPPI of the parameter subset $\bm\phi$, as follows:

\begin{align*}\mbox{EVPPI}(\bm\phi) = &\ \mbox{E}_{\bm\phi} \left[\max_t \mbox{E}_{\bm\psi\mid\bm\phi} \left[\mbox{NB}_t(\bm\theta) \right]\right]- \max_t \mbox{E}_{\bm\theta}\left[ \mbox{NB}_t(\bm\theta)\right] \\
= &\ \mbox{E}_{\bm\xi} \left[\mbox{E}_{\bm\xi^c\mid\bm\xi} \left[\max_t \mbox{E}_{\bm\psi\mid\bm\phi} \left[\mbox{NB}_t(\bm\theta) \right]\right]\right]- \max_t \mbox{E}_{\bm\theta}\left[ \mbox{NB}_t(\bm\theta)\right] \\
\geq &\ \mbox{E}_{\bm\xi} \left[\max_t \mbox{E}_{\bm\xi^c\mid\bm\xi} \left[\mbox{E}_{\bm\psi\mid\bm\phi} \left[\mbox{NB}_t(\bm\theta) \right]\right]\right]- \max_t \mbox{E}_{\bm\theta}\left[ \mbox{NB}_t(\bm\theta)\right] \\
&\ \mbox{E}_{\bm\xi} \left[\max_t \mbox{E}_{\bm\xi^c\mid\bm\xi} \left[\mbox{E}_{\bm\psi\mid(\bm\xi,\bm\xi^c)} \left[\mbox{NB}_t(\bm\theta) \right]\right]\right]- \max_t \mbox{E}_{\bm\theta}\left[ \mbox{NB}_t(\bm\theta)\right] \\
=&\ \mbox{E}_{\bm\xi} \left[\max_t \mbox{E}_{(\bm\psi,\bm\xi^c)\mid\bm\xi} \left[\mbox{NB}_t(\bm\theta) \right]\right]- \max_t \mbox{E}_{\bm\theta}\left[ \mbox{NB}_t(\bm\theta)\right] = \mbox{EVPPI}(\bm\xi)\\
\end{align*}
by Jensen's inequality as the function $\max(\cdot)$ is  convex.

\end{document}